# Microbial assessment in a rare Norwegian book collection: a One Health approach to cultural heritage

Sílvia O. Sequeira[1,2]*; Ekaterina Pasnak[1]; Carla Viegas[3,4]*; Bianca Gomes[3,5]; Marta Dias[3,4]; Renata Cervantes[3,4]; Pedro Pena[3,4]; Magdalena Twarużek[6]; Robert Kosicki[6]; Susana Viegas[3,4]; Liliana Aranha Caetano[3,7]; Maria João Penetra[8], Inês Santos[8,9], Ana Teresa Caldeira[8,9], Catarina Pinheiro [2,8,9]*

1 LAQV-REQUIMTE, Department of Conservation and Restoration, NOVA School of Sciences and Technology, Nova University of Lisbon, Campus da Caparica, 2829-516 Caparica, Portugal
2 Laboratório José de Figueiredo, Museus e Monumentos de Portugal, Rua das Janelas Verdes, 1249-018 Lisboa, Portugal
3 H&TRC – Health & Technology Research Center, ESTeSL – Escola Superior de Tecnologia e Saúde, Instituto Politécnico de Lisboa; 1990-096 Lisbon, Portugal;
4 NOVA National School of Public Health, Public Health Research Centre, Comprehensive Health Research Center, CHRC, NOVA University Lisbon, 1600-560 Lisbon, Portugal;
5 CE3C—Center for Ecology, Evolution and Environmental Change, Faculdade de Ciências, Universidade de Lisboa, 1749-016 Lisbon, Portugal;
6 Kazimierz Wielki University, Faculty of Biological Sciences, Department of Physiology and Toxicology, Chodkiewicza 30, 85–064 Bydgoszcz, Poland;
7 Research Institute for Medicines (iMed.uLisboa), Faculty of Pharmacy, University of Lisbon, 1649-003 Lisbon, Portugal;
8 HERCULES Laboratory, Évora University, Palácio do Vimioso, Largo Marquês de Marialva 8, 7000-809 Évora, Portugal
9 IN2PAST — Associate Laboratory for Research and Innovation in Heritage, Arts, Sustainability and Territory, University of Évora, Largo Marquês de Marialva 8, 7000-809 Évora, Portugal.

*corresponding authors: silvia.sequeira@museusemonumentos.pt; carla.viegas@estesl.ipl.pt; anapinheiro@museusemonumentos.pt

## Abstract

Microbial contamination poses a threat to both the preservation of library and archival collections and the health of staff and users. This study investigated the microbial communities and potential health risks associated with the UNESCO-classified Norwegian Sea Trade Archive (NSTA) collection exhibiting visible microbial colonization and staff health concerns. Dust samples from book surfaces and the storage environment were analysed using culturing methods, qPCR, Next Generation Sequencing, and mycotoxin, cytotoxicity and azole resistance assays. *Penicillium* sp., *Aspergillus* sp., and *Cladosporium* sp. were the most common fungi identified, with some potentially toxic species like *Stachybotrys* sp., *Toxicladosporium* sp. and *Aspergillus* section *Fumigati.* Fungal resistance to azoles was not detected. Only one mycotoxin, sterigmatocystin, was found in a heavily contaminated book. Dust extracts from books exhibited moderate to high cytotoxicity on human lung cells, suggesting a potential respiratory risk. The collection had higher contamination levels compared to the storage environment, likely due to improved storage conditions. Even though, overall low contamination levels were obtained, which might be underestimated due to the presence of salt (from cod preservation) that could have interfered with the analyses. This study underlines the importance of monitoring microbial communities and implementing proper storage measures to safeguard cultural heritage and staff well-being.

## 1. Introduction

Collections in libraries and archives predominantly consist of paper materials, which due to their organic composition and hygroscopic behaviour are prone to microbial colonization. By degrading paper-based materials, microorganisms can also contribute to the distinctive musty odour often associated with historical libraries [1], [2]. To mitigate the growth of mould, libraries and archives implement strategies such as maintaining efficient ventilation and keeping relative humidity levels below 60%. However, challenges arise when climate control systems fail or cannot adequately handle sudden spikes in temperature and humidity. Emergencies like leaks or floods can also lead to microbial outbreaks and even if the immediate issue is resolved, the environment stabilized, and collections cleaned, fungal spores may still linger within the paper fibres. Several studies have so far identified causative agents for microbial biodeterioration on paper, but not considering the health impact of such contamination [3]. Under sound environmental conditions and for immunocompetent hosts, most fungi are harmless. However, 19% of the species so far identified in libraries and archives can cause various health effects [4]. Besides spores and fungal remains, such as mycelia, which can cause allergic reactions, fungi also excrete exotoxins during their growth – mycotoxins – which can cause allergies, asthma other health-related issues among staff and employers [3], [5], [6], [7].

The Norwegian Sea Trade Archive (NSTA), housed in the University Library of Bergen, Norway (ULB) documents the activity of private companies that traded stock and salted dry cod fish from the 16th until the middle of the 20th century [8]. This unique collection, composed of 2311 items, mostly accounting books, is included in UNESCO´s Memory of the World Register due to its cultural, historical, and economic significance. Historically, the fish was stored on the ground floor of the wooden buildings located along the quay at Bergen Port while the accounting was done on the first floor. Because heating the building was prohibited due to fear of fire, the environment was humid and cold leading to biological deterioration of the collection at that time. Adding to this, in 2016 the HVAC system in the NSTA-ULB storage where this collection was housed was out of order and the environment became again ideal for microbial development. The books also exhibit soiling and a strong codfish smell. In the last decades, library workers have been reporting various skin, eye, and respiratory symptoms from contact with the collection – symptoms that could be associated with occupational exposure to chemical contaminants but also to moulds, mycotoxins, and endotoxins (toxins of bacterial origin). As a result, access to this important collection is currently very restricted.

Aiming to understand what the cause of the manifested symptoms could be and characterize the microbial contamination around and inside the books of this important collection, we have joined exposure science and cultural heritage conservation, in a One Health approach [9]. Using passive sampling, culture-based methods, and molecular tools, we aimed to identify and quantify the microorganisms in both the books and the storage environment, evaluate the resistance of identified species to azole-based fungicides, assess the presence of mycotoxins, and determine the potential health effects due to exposure through cytotoxicity analyses.

## 2. Material and methods

### 2.1 Sampling campaign

Sampling of the books was performed by vacuuming the surface of the pages and covers with a museum-grade vacuum cleaner (Muntz museum vacuum cleaner 555 MU with HEPA filter), having 8 cm squares of filter paper (coffee filters nº 4, Auchan, France) between the hose and the nozzle to capture the aspirated particles (FP samples). Before sampling, each filter paper was sterilized under UV radiation for 1 hour inside a biological safety cabinet and kept in sterile bags [10]. Sterilized coffee filter pouches, uncut, were also used to wrap the paper squares after sampling so that the collected dust would not electrostatically adhere to the sterile plastic bags. Eight books in total (n=8) were sampled: four books showing clear signs of fungal colonization and another four with no visual evidence of such deterioration. From each visually fungal-affected book (n=4), two samples were collected, for comparison: a first sample, focusing on the localized fungal colonies (samples FP×A – where × is the sample number), and a second sampling encompassing the rest of the book (samples FPxB), which summed up to a total of 12 samples collected from the books. With the vacuum still on, so that the dust would not fall off, the nozzle was disconnected, and the filter paper sample was put inside the coffee filter pouch (with tweezers) and then placed inside a sterile bag. Both the vacuum cleaner hose and the tweezers were disinfected with 70% ethanol between each sample collection. One filter paper inside the coffee filter pouch was left unused and served as a control sample.

To analyse the storage environment, electrostatic dust collectors (EDCs), which are simple electrostatic cloths (Swiffer), were used to perform a passive sampling [10]. The EDCs, cut to 95 × 130 mm, were taped (Scotch tape) to the interior of a printing paper bifolio (210 ×145 mm). The prepared EDCs samplers were sterilized the same way as the filter papers, by 1 hour UV exposure and stored in sterilized bags until application. 11 EDCs were placed in the storage room, and one was kept in the sterile bag and used as a control.

The EDCs were distributed uniformly over the storage area, placed open over shelves or cabinets, at c. 1.5 m high, to collect airborne dust, which is the one that contributes to human exposure by inhalation. They were secured in place with scotch tape to prevent accidental moving and a warning sign "Do not touch" (in Norwegian) was placed next to them. EDCs were maintained in place for 31 days and then collected into individual sterile bags and shipped for analysis. The collection of the EDCs involved only the folding of the paper bifolio, avoiding direct contact with the interior.

All samples were analysed according to the diagram presented in Figure 1.

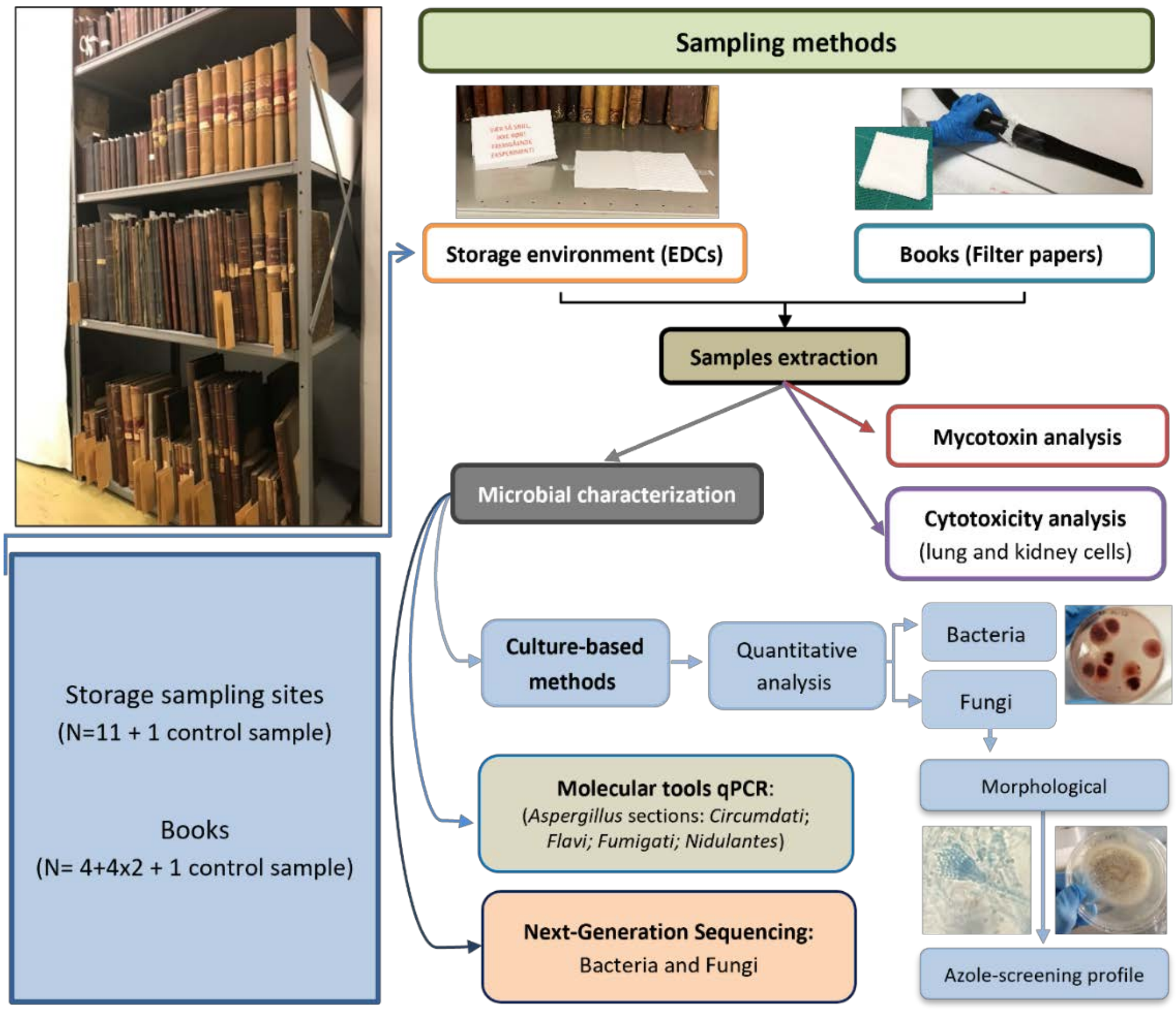

*Figure 1. Diagram of employed analyses for each type of sample.*

## 2.2 Culture-based methods

All samples were extracted and further analysed as previously described [11]. Briefly, they were washed with 0.1% Tween 80 saline (0.9%NaCl) solution (250 rpm, 30 min), as follows: 10 mL solution for the collected dust from the filter papers (books aspiration) (2 $cm^2$); 20 ml solution for each EDC (environment). The obtained extracts were plated (150 ml) in selective culture media for fungi, namely: malt extract agar (MEA) supplemented with chloramphenicol (0.05%), dichloran-glycerol agar (DG18), and bacteria: tryptic soy agar (TSA) supplemented with nystatin (0.2%), and violet red bile agar (VRBA) and incubated at optimal temperature and time conditions for fungi and bacteria [12]. Fungal species/sections were identified microscopically through macro and microscopic characteristics as noted by De Hoog [13].

Filter papers and EDCs extracts were screened for antifungal resistance towards four antifungal agents at set concentrations in Sabouraud agar medium (SDA): 4.0 mg/L itraconazole (ICZ), 2.0 mg/L voriconazole (VCZ) and 0.5 mg/L posaconazole (PCZ), at 27°C for 48h, as previously reported [12].

The same extracts were also employed for molecular detection of the selected fungal targets (*Aspergillus* sections *Circumdati*, *Flavi*, *Fumigati* and *Nidulantes*) following the previously published procedures [12]. Fungal DNA was extracted using the ZR Fungal/Bacterial DNA MiniPrep Kit (Zymo Research, Irvine, USA) following the producer's instructions, and molecular identification was accomplished by Real-Time PCR (qPCR) using the CFX-Connect PCR System (Bio-Rad). For each amplified gene, a non-template control and a positive control (DNA obtained from reference strains kindly provided by the Mycology laboratory of the National Institute of Health Dr Ricardo Jorge) were employed.

## 2.3 Identification of microorganisms by NGS

The extracted DNA was used also to perform environmental metagenomics. The DNA concentration was determined by fluorometry with Quantus™ Fluorometer ONE dsDNA quantification kit (Promega, Madison, USA), according to the manufacturer's instructions. Microbial communities were characterized according to published protocols [14], [14], [15]. Bacterial communities were characterized by Illumina Sequencing technology for the 16S rRNA V3-V4 region. Metagenomic DNA was amplified for the hypervariable regions with specific primers and further re-amplified in a limited-cycle PCR reaction to attach a sequencing adaptor and dual indexes. The prokaryotic population was characterized using the 16S V3 forward primer 341F 5'-CCTACGGGNGGCWGCAG-3' and 16S V4 reverse primer 805R 5'-GACTACHVGGGTATCTAATCC-3' [16], [17]. Besides, the 16S target-specific sequences, the primers also contained adaptor sequences allowing uniform amplification of the library with high complexity ready for downstream NGS sequencing on Illumina Miseq. The hypervariable regions were amplified for each sample by PCR, in a LifeEco Thermal Cycler (Bioer Technology, China), for a total volume of 25 µL, containing 10 µL of Bioline My Taq HS Mix, 5 µL of each primer (1 mM) and 5 µL of DNA. The PCR program consisted of 1 min of denaturation at 95°C, followed by 35 cycles of denaturation at 95°C for 15 seconds, annealing at 55°C for 15 seconds and polymerization at 72°C for 10 seconds, and a final extension at 72°C for 2 min. For the eucaryotic communities, the ITS3 of the nuclear ribosomal RNA genes was amplified using the following primers: ITS3_1F 5'- CATCGATGAAGAACGCAG-3', ITS3_2F 5'-CAACGATGAAGAACGCAG-3', ITS3_3F 5'- CACCGATGAAGAACGCAG -3', ITS3_4F 5'-CATCGATGAAGAACGTAG 3', ITS3_5F 5'-CATCGATGAAGAACGTGG-3', ITS3_10F (5'-CATCGATGAAGAACGCTG-3', ITS3_001R 5'- TCCTSCGCTTATTGATATGC -3'. The hypervariable regions were amplified for each sample by PCR, in a LifeEco Thermal Cycler (Bioer Technology, China), for a total volume of 25 µL, containing 10µL of Bioline My Taq HS Mix, 3.5µL of pool primer (10 mM) and 5µL of DNA. The PCR program consisted of 2 min of denaturation at 95°C, followed by 35 cycles of denaturation at 95°C for 30 seconds, annealing at 55°C for 30 seconds and polymerization at 72°C for 20 seconds, and a final extension at 72°C for 2 min. Negative controls were included for all amplification reactions. The amplification products were detected by electrophoresis in a 2% (w/v) agarose gel with a 100 bp DNA ladder and the gel was stained with Green Premium and visualized under UV light in a Bio-Rad Molecular

Imager ® Gel Doc™ XR+ Imaging System. The amplified fragments were purified using the High Prep TM PCR Cleanup System according to the manufacturer's instructions. Next, dual indexes and Illumina sequencing adapters were attached to both ends using the Illumina Nextera XT Index Kit (Illumina, San Diego, CA, USA), using 25 µL of 2X KAPA HiFi HotStart Ready Mix, 10 µL of H 2 O RNase Free, 5 µL of Illumina Nextera XT Index Primers 1 (N7XX), 5 µL of Illumina Nextera XT Index Primers 2 (N5XX) and 5 µL of amplicon PCR product purified, for a total of 50 µL. The PCR index program consisted of a 3-minute denaturation step at 95°C, followed by 8 cycles of amplification: denaturation at 95°C for 30 seconds, annealing at 55°C for 30 seconds, and polymerization at 72°C for 30 seconds, and a final extension at 72°C for 5 minutes. The metagenomic libraries/ Index PCR products were detected by electrophoresis in a 2% (w/v) agarose gel with a 100 bp DNA ladder. The amplicon products were subsequently purified using the HighPrep™ PCR Cleanup System, according to the manufacturer's instructions. The library concentration was determined by fluorometry with Quantus™ Fluorometer ONE dsDNA quantification kit (Promega, Madison, USA), according to the manufacturer's instructions. Libraries were normalized and pooled to 4 nM. Pooled libraries were denatured and diluted to a final concentration of 10 pM with a 15% PhiX (Illumina) control. Sequencing was performed using the MiSeq Reagent Nano Kit V2 in the Illumina MiSeq System. Samples sequencing was performed using a 2 × 250 paired-end (PE)configuration; image analysis and base calling were conducted by the MiSeq Control Software (MCS) directly on the MiSeq instrument (Illumina, San Diego, CA, USA). The forward and reverse reads were merged by overlapping paired-end reads using the AdapterRemoval v2.1.5 [18] software with default parameters. The QIIME package v1.8.0 [19] was used for Operational Taxonomic Units (OTU) generation, taxonomic identification, and sample diversity and richness indexes calculation. Sample IDs were assigned to the merged reads and converted to fasta format (split_libraries_fastq.py, QIIME). Chimeric merged reads were detected and removed using UCHIME [20] against the Greengenes v13.8 database [21] for V3-V4 samples. OTUs were selected at a 97% similarity threshold using the open reference strategy. First, merged reads were pre-filtered by removing sequences with a similarity lower than 60% against Greengenes v13.8 databases. The remaining merged reads were then clustered at 97% similarity against the same databases listed above. Merged reads that did not cluster in the previous step were again clustered inOTU at 97% similarity. OTUs with less than two reads were removed from the OTU table. A representative sequence of each OTU was then selected for taxonomy assignment.

## 2.4 Analysis of mycotoxins

Sample preparation and chromatographic analysis of mycotoxins followed the procedure outlined in Viegas et al. [22]. In summary, 0.10 g of dust collected from the books were subjected to vigorous shaking for 60 minutes, using 3.0 ml of an acetonitrile/water/acetic acid mixture (79/20/1; v/v/v). After 5-minute centrifugation at 5000 rpm, 2ml of the extract was evaporated to dryness under a stream of nitrogen and then reconstituted in a 400 µl of methanol/water mixture (2/8; v/v) and centrifuged again for 30 min at 14500 rpm. Thus, the sample dilution factor was 6.

The detection of mycotoxins was conducted using a high-performance liquid chromatograph (HPLC) system, specifically the Nexera model from Shimadzu (Kyoto, Japan), coupled with a mass spectrometry detector, the 5500 QTrap from Sciex (Foster City, USA). Mycotoxins were separated by chromatography on a Gemini C18 column (150 × 4.6 mm, 5 µm) manufactured by Phenomenex in Torrance, CA, USA. The flow rate was set at 1 ml/min, and a 5 µL injection volume was employed.

Two distinct mobile phases were utilized: Phase A, comprising methanol/water/acetic acid in a ratio of 10/89/1 (v/v/v), and Phase B, consisting of methanol/water/acetic acid in a ratio of 97/2/1 (v/v/v). Both mobile phases were supplemented with 5 mmol/L of ammonium acetate. The chromatographic gradient proceeded as follows: initial elution with 0% B up to 2.0 minutes, followed by a linear increase to 50% B from 2.0 to 5.0 minutes, further ramping up to 100% B from 5.0 to 14.0 minutes, maintaining 100% B until 18.0 minutes, and ultimately returning to the initial 0% B composition by 22.5 minutes.

Tandem mass spectrometry analysis was conducted in the scheduled multiple reaction monitoring (sMRM) mode for both negative and positive polarities within a single chromatographic run. The electrospray ionization (ESI) source parameters were set as follows: a curtain gas at 30 psi, collision gas at a medium level, ion spray voltage at -4500 V (negative polarity) and 5500 V (positive polarity), ion source temperature maintained at 550°C, ion source gas1 at 80 psi, and ion source gas2 at 80 psi. Table S1 (Supplementary material) shows the instrument settings optimized for product ions of each compound. The Analyst 1.6.2 software (Sciex, Foster City, CA) was used for data acquisition and processing.

## 2.5 Analysis of cytotoxicity

Cytotoxicity was measured by using the MTT (3-(4,5-dimethylthiazol-2-yl)-2,5-diphenyltetrazolium bromide) test on swine kidney (SK) and human lung epithelial (A549 cells. For that purpose, each settled dust sample (N=12) and EDC sample (N=12) was shaken with 5 mL ACN/H2O (84/16, v/v) for 30 min, centrifuged, and the supernatant was taken and evaporated to dryness under a gentle stream of nitrogen at 40 °C. Evaporated extracts were dissolved in 1 mL of a mixture of ethanol, dimethyl sulfoxide, and Minimum Essential Medium Eagle (MEM) (1.7 + 0.3 + 98,v/v/v). Samples were tested in varying concentrations using the 2-fold serial dilution method.

Cells were seeded on a 96-well microtiter plate and incubated with 100 µl of the prepared sample dilutions per well for 48 hours at 37 °C in a humidified atmosphere with 5% CO2. Subsequently, MTT (3-(4,5-dimethylthiazol-2-yl)-2,5-diphenyltetrazolium bromide) solution (20 µL) was added, and plates were incubated for another 4 h. The supernatant was then removed, and 100 µL dimethyl sulfoxide (DMSO) was added to each well. The formation of formazan was measured by spectrophotometric absorbance using an ELISA microplate reader (ELISA LEDETECT 96, Biomed Dr Wieser GmbH, Salzburg, Austria) at a wavelength of 510 nm (=maximum absorption wavelength of formazan derivatives). The lowest sample concentration dropping absorption to <50% of cell metabolic activity (IC50) was defined as the threshold toxicity level.

# 3. Results

## 3.1 Microbial characterisation

### *3.1.1 Culturing Media: Bacteria and Fungi*

Regarding bacterial contamination, filter papers (vacuumed books) presented the highest total value on total bacteria (TSA: $9.69\times10^{3}$ CFU.m$^{-2}$), when compared to Gram-negative bacteria (VRBA: $2.81\times10^{2}$ CFU.m$^{-2}$). The same was seen in EDC samples, where the highest total value was presented on total bacteria (TSA: $2.87\times10^{2}$ CFU.m-2.day$^{-1}$), followed by Gram-negative bacteria (VRBA: $1.06\times10^{1}$ CFU.m$^{-2}$.day$^{-1}$) (Figure 2).

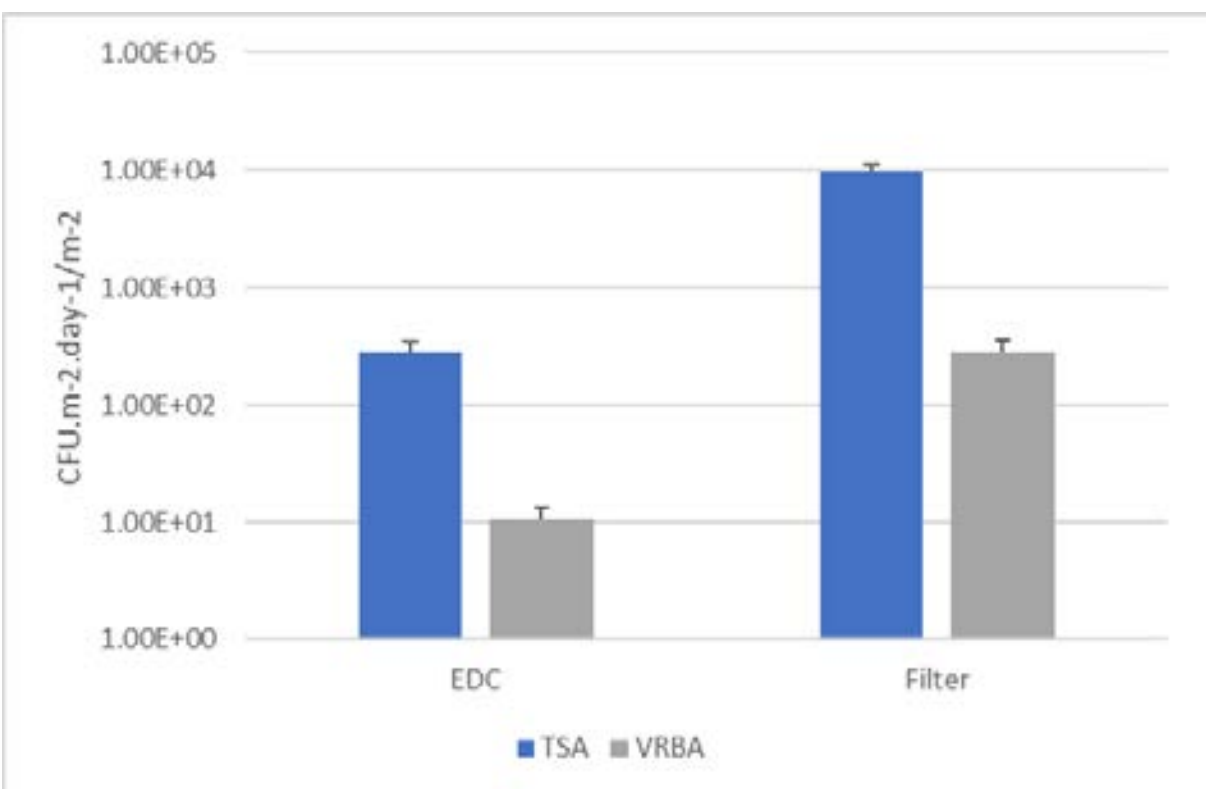

*Figure 2. Total bacteria (TSA) and Gram-negative (VRBA) median values from EDCs (CFU.m-2.day-1) and filters (CFU.m-2).*

On fungal contamination, filter papers presented the highest value on DG18 ($2.5x10^2$ $CFU.m^{-2}$), when compared to MEA ($1.88x10^2$ $CFU.m^{-2}$), while on EDC samples, MEA had the highest value ($6.24x10^1$ CFU.m-2.$day^{-1}$), followed by DG18 ($1.87x10^1$ CFU.m-2.$day^{-1}$) (Figure 3).

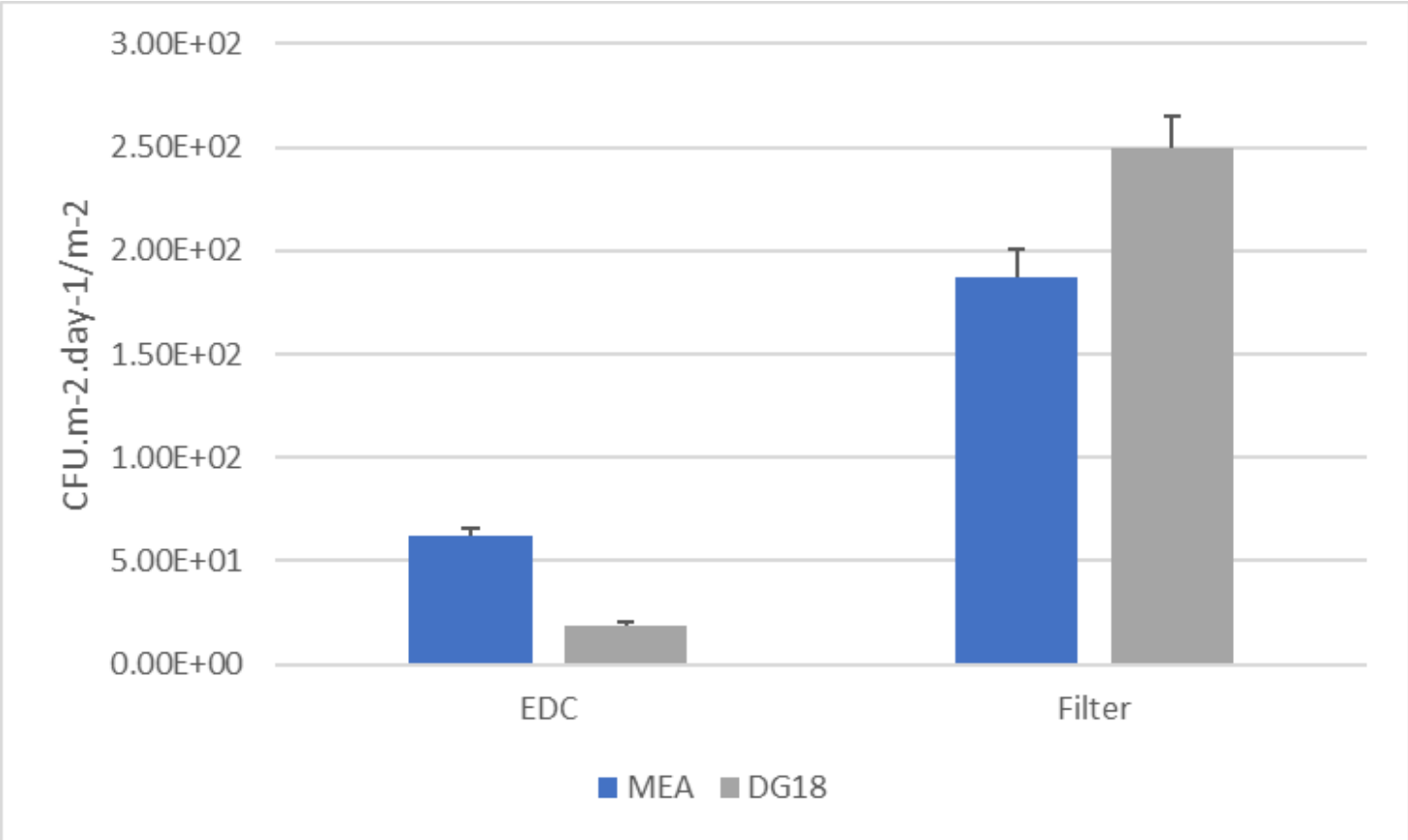

*Figure 3. Fungal contamination (MEA and DG18) median values from EDC (CFU.m-2.$day^{-1}$) and filters ($CFU.m^{-2}$).*

There was a prevalence of *Penicillium* sp. followed by *Cladosporium* sp. in both EDCs and filter papers (MEA culture medium). Using the lower water activity medium (DG18), there was a higher prevalence of *Cladosporium* sp. on EDCs, whereas in filter papers *Aspergillus* sp. was the most prevalent genera (Table 1).

In the EDCs, only two *Aspergillus* sections were identified on MEA – *Nigri* and *Nidulantes* (50 %) – and only *Nidulantes* appeared on DG18. In the filter paper samples, the *Fumigati* (67 %) section was also identified on DG18 alongside the *Nigri* section (33 %) (Figure 4).

Concerning antifungal resistance, although *Penicillium* sp. (9.37x100 $CFU.m^{-2}$.day) and *Cladosporium* sp. (3.12x100 $CFU.m^{-2}$.day) from EDCs, and *Aspergillus* section *Nidulantes* (3.12x100 $CFU.m^{-2}$.day) from filter papers were able to grow on the control SDA plates, no fungal growth was observed in the media supplemented with the four antifungal agents. The *Aspergillus* sections targeted by qPCR were not detected in the analysed samples.

*Table 1. Fungal distribution on MEA and DG18 from EDC and vacuum filter samples.*

| | MEA | | | DG18 | | |
|---|---|---|---|---|---|---|
| **Samples** | **Genera** | **$CFU.m^{-2}.day^{-1}/m^{-2}$** | **%** | **Species** | **$CFU.m^{-2}.day^{-1}/m^{-2}$** | **%** |
| **EDCs** | *Penicillium* sp. | 1.25E+01 | 40 | *Cladosporium* sp. | 6.24E+00 | 67 |
| | *Cladosporium* sp. | 9.37E+00 | 30 | *Aspergillus* sp. | 3.12E+00 | 33 |
| | *Aspergillus* sp. | 6.24E+00 | 20 | | | |
| | *Aureobasidium* sp. | 3.12E+00 | 10 | | | |
| **Paper filters** | *Penicillium* sp. | 6.25E+01 | 67 | *Aspergillus* sp. | 9.38E+01 | 75 |
| | *Cladosporium* sp. | 3.13E+01 | 33 | *Penicillium* sp. | 3.13E+01 | 25 |

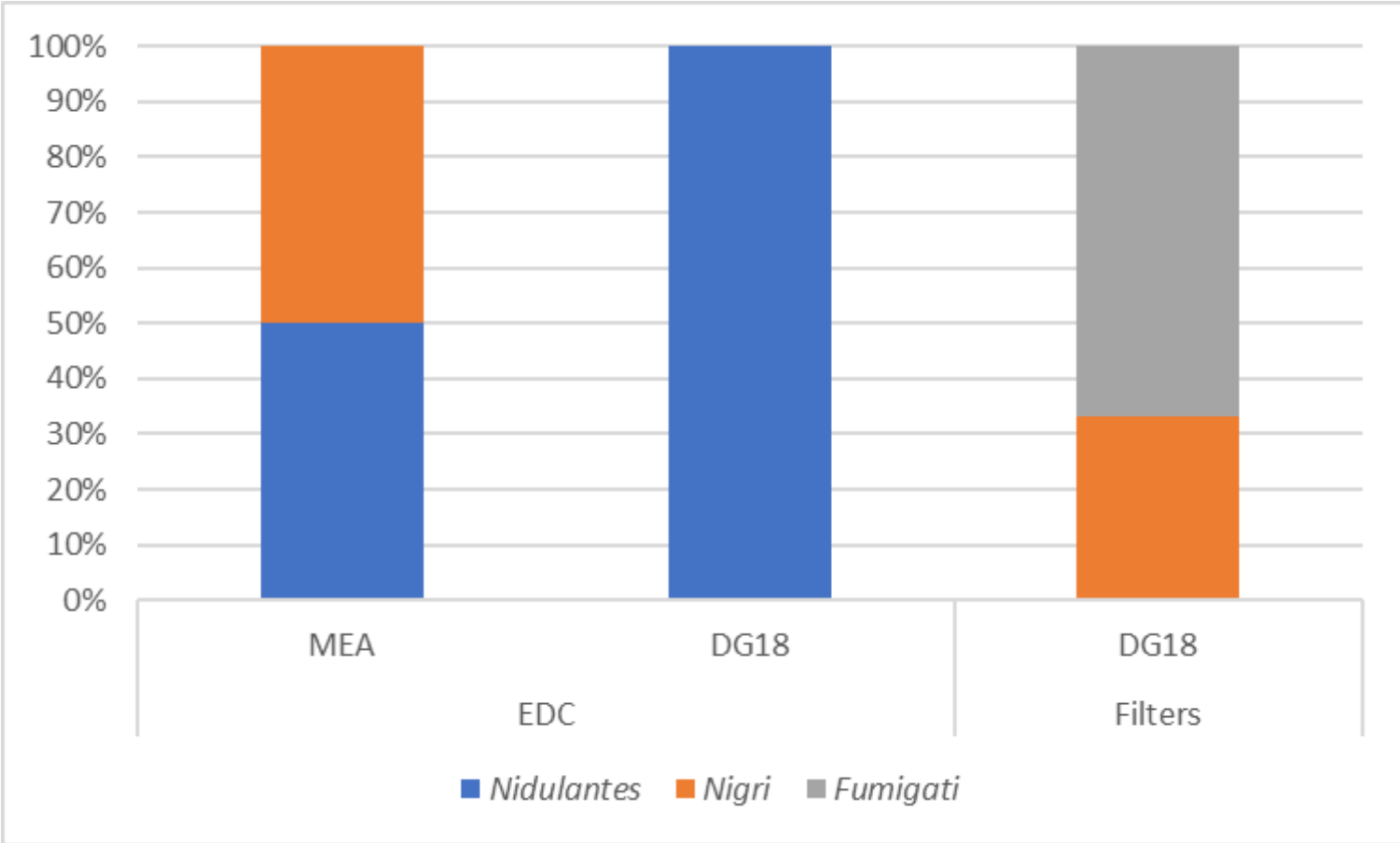

*Figure 4. Aspergillus sections distribution on MEA and DG18 from EDC and paper filters. No* Aspergillus *section was detected on MEA for the paper filters.*

### *3.1.2 Next Generation Sequencing: Bacteria and Fungi*

Despite normal levels of extracted DNA, quantified in the pre-treatment phase, low amplification rates were obtained. The number of OTUs for Procaryota (bacteria) in all samples was 553 and for Eucaryotes (fungi) it was 67, and their distribution is presented in Figures 5 and 6, respectively.

The procaryotic diversity profile obtained for all samples was very similar and the number of reads obtained was both low and very similar amongst the samples and when comparing the samples (EDC1 to EDC12; PF 1 to PF8B) to the respective controls (Figure 5). Nevertheless, the profile delivered a predominance of Actinobacteria, Proteobacteria and Cyanobacteria in all samples followed by the Firmicutes and other less represented phyla. Within these, it is possible to identify Euryarchaeota, or Archae, known for their ability to thrive in extreme environments such as heavily salted ones [23]. All sequences regarding the Procariota were deposited in NCBI under Bioproject PRJNA1071534.

Regarding Eucaryota, a sharp presence of Actinobacteria followed by Basidiomycota is observed in all samples (Figure 6). Unlike the results obtained for the Procaryota, however, one Eucaryotic sample – PF6B – showed a dissimilar result, substantially different from the controls and the remaining samples (Figure 7). It was the case of a heavily contaminated book, in which the Ascomycota phylum accounts for 98% of the contamination. In terms of Class distribution, Dothidomycetes account for 43%, followed by Eurotiomycetes (20%), Saccharomycetes (17%) and Sordariomycetes (11.4%). The Order distribution presents Botryosphaeriales (20%) followed by Eurotyales (19%), Capnodiales (18%), Saccharomycetales (17%) and Hypocreales (13%).

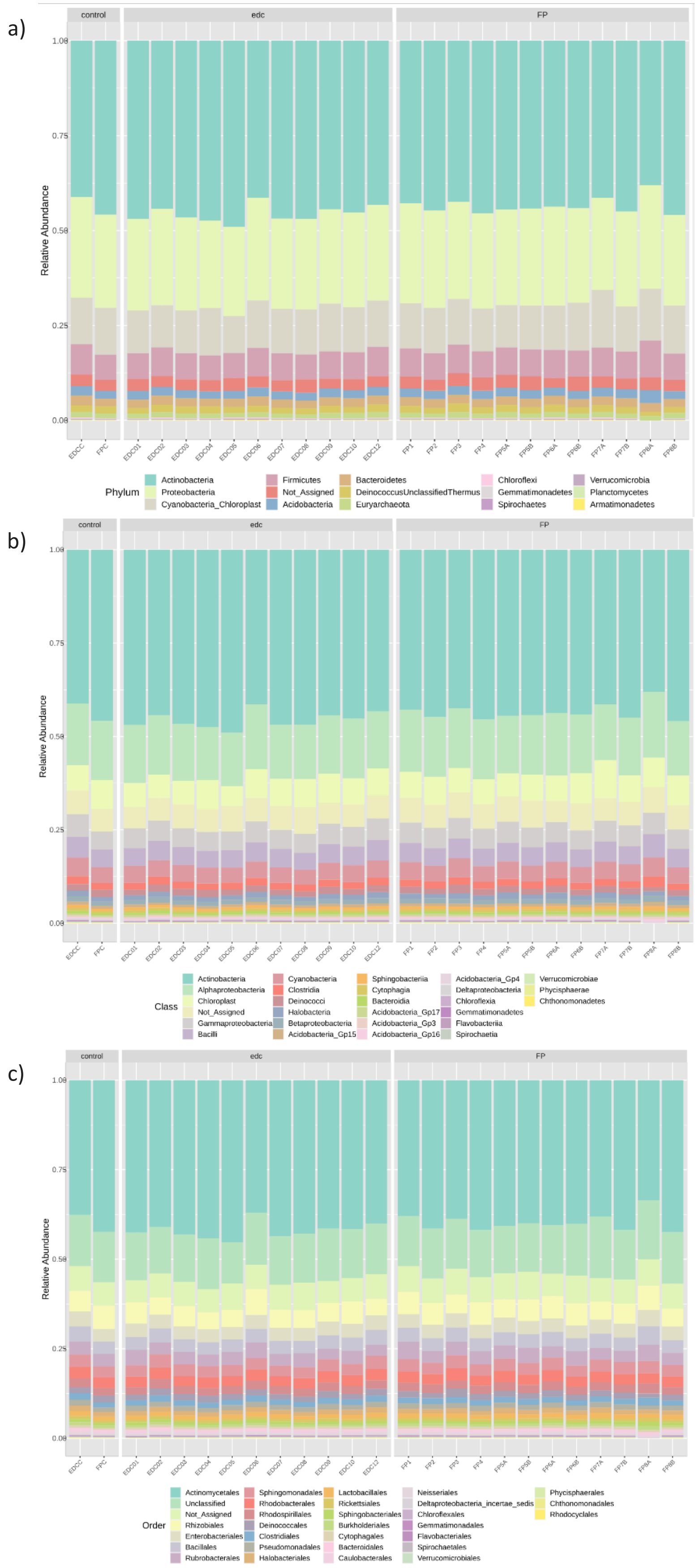

*Figure 5. Procaryota composition of the controls, EDCs and paper filters. The bar plots show the relative abundances (%) at phylum (a), class (b) and order levels (c). C- controls; EDCs – Electrostatic dust cloths; FP – Filter papers.*

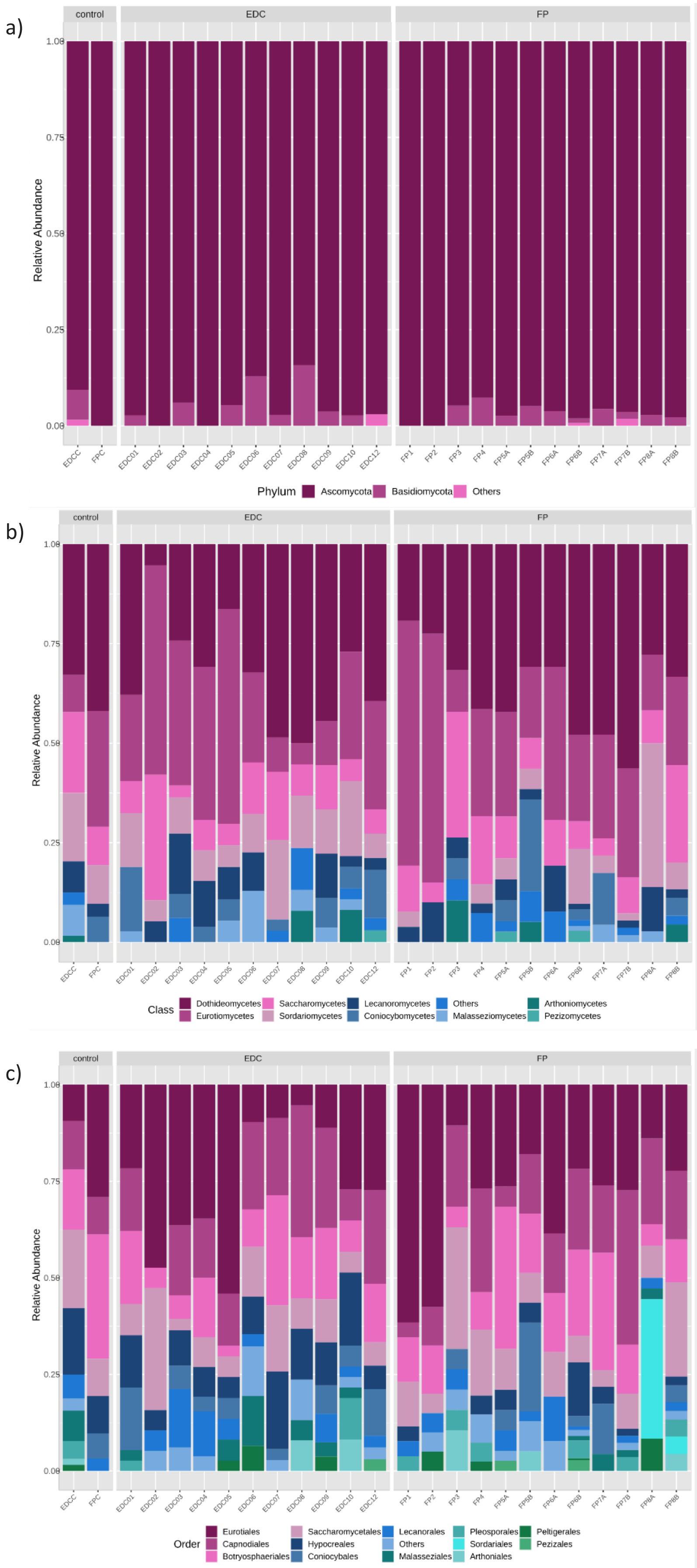

*Figure 6. Eucaryota composition of the controls, EDCs and* Filter papers*. The bar plots show the relative abundances (%) at phylum* (a)*, class* (b) *and order* (c) *levels. C- controls; EDCs – Electrostatic dust clothes; FP – Filter papers.*

When analysing Figure 7 it is possible to verify the presence of relevant genera in sample VF6B such as *Stachybotrys sp.*, *Toxicocladosporium* sp. and *Aspergillus*, sp., all of them with possible health implications [24], [25]. The remaining samples (controls included) present a somewhat similar distribution (see figures 6 and 7) but the possibility that this could be the result of cross-contamination from sample FP6B cannot be discarded because this method is very sensitive and amplifies minute amounts of DNA through several cycles of amplification. So, despite careful manipulation of the samples, more research is needed to confirm this hypothesis.

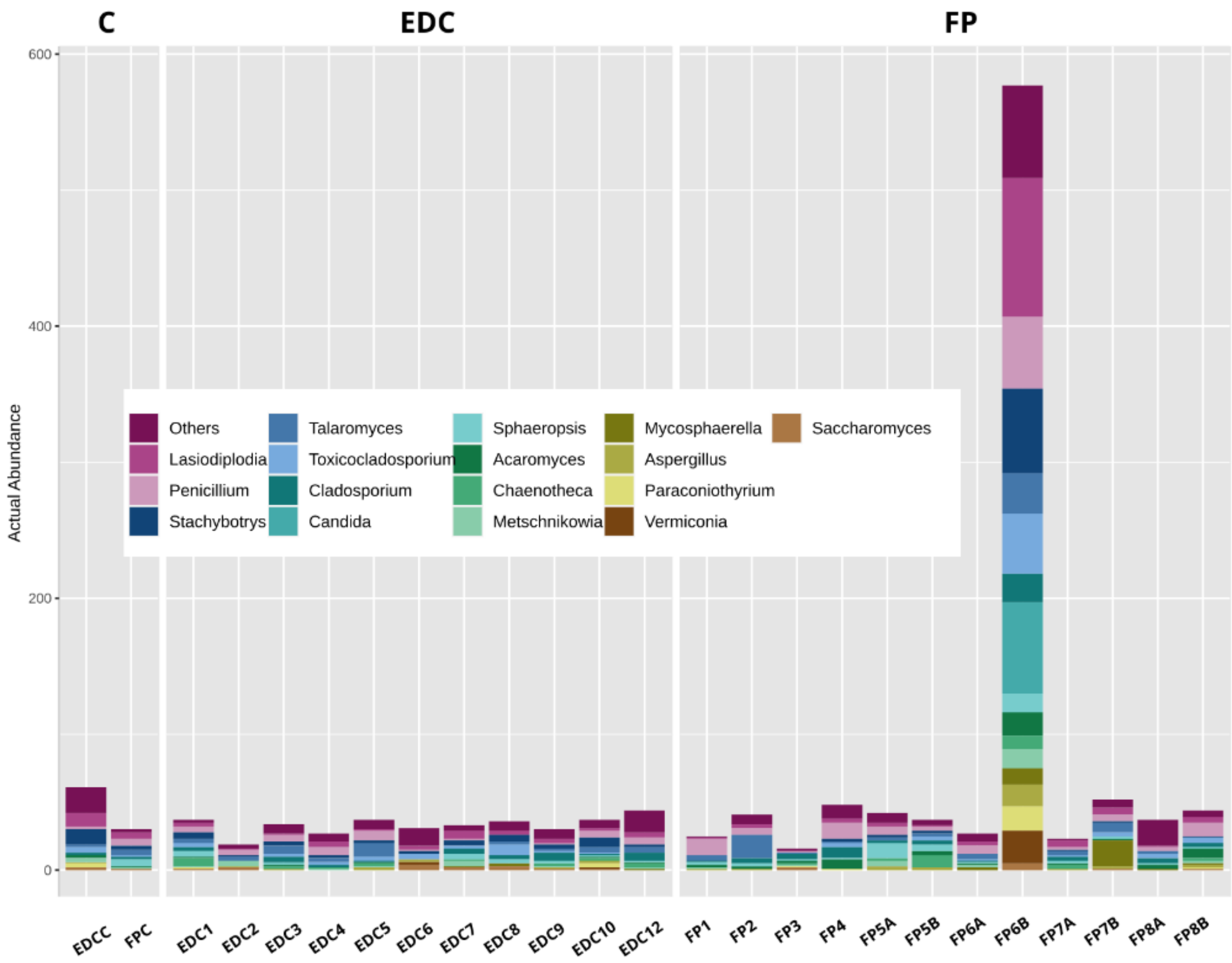

*Figure 7. Relative abundance of fungal genera in all samples. The lowest abundances were grouped under the name "Others".*

## 3.3 Mycotoxins

Only one vacuum filter sample (FP6B – the same one showing a different fungal profile) presented positive results. The mycotoxin detected was sterigmatocystin (<LOD 2.2 ng). None of the EDC samples presented detectable contamination.

## 3.4 Cytotoxicity

The effect of vacuumed dust from books and EDCs' contaminants on cell viability was assessed by employing the MTT test on swine kidney (SK) and human lung epithelial (A549) cells (Table 2). These cells are relevant *in vitro* models for toxicological assessment of human exposure to biological contaminants [26], [27].

In A549 cells, filter papers ranged from low (25%) to high (42%) cytotoxicity, whereas EDCs showed low cytotoxicity (25%) only. No cytotoxicity was observed in swine kidney cells for any sampling method.

*Table 2. Distribution of threshold toxicity (IC50) among human lung epithelial (A549) and swine kidney (SK) cells on EDCs and filter paper (FP) samples.*

| | | A549 cells | | SK cells | |
|---|---|---|---|---|---|
| **Cytotoxicity level** | **Sample dilution** | **FP** | **EDC** | **FP** | **EDC** |
| **Low** | **1:2** | 25% | 25% | 0% | 0% |
| **Moderate** | **1:4** | 8% | 0% | 0% | 0% |
| | **1:8** | 25% | 0% | 0% | 0% |
| **High** | **1:16** | 17% | 0% | 0% | 0% |
| | **1:32** | 25% | 0% | 0% | 0% |

## 4. Discussion

The sampling and analytical techniques proposed in this study have been successfully tested by our team in settings where microbiological contamination needed to be addressed for its possible impacts on human health [28], [29]. It was, therefore, the first choice when assessing a setting where undiagnosed ailments keep affecting not only the well-being of the conservators-restorers that handle this collection but also pose a risk to the general public who wishes to consult this historic documentation. The collection – or part of it – could, therefore, be seen as an itinerant hazard as symptoms seem to appear whenever it is manipulated.

The fact that part of this collection is visually contaminated by microorganisms sustained the hypothesis of a causative biological agent. The results, however, did not confirm a high burden of contamination nor a great biological diversity in either the environment (EDCs) or the filter papers containing the particulate collected from the books. This was true for both the traditional culturing approaches and the DNA/PCR-based methodologies.

When looking for reasons for this occurrence – especially when contamination is so visible in the books - one cannot avoid wondering if the environment itself could have played a role since all the sampled documentation (and possibly the environment on which it rests) might still contain salt, a permanent presence wherever cod is preserved. This being the case, then the microbial communities are probably adapted to high osmotic pressure and salty environments and conventional analytical protocols do not consider specific environments because these tend to be rare and specific. By not supplementing the culture media with NaCl and by not foreseeing its presence as a possible contamination agent for DNA-based strategies, we may have missed relevant information. Thus, the obtained results may be underestimated.

DNA is ubiquitous in DNA extraction kits and other laboratory reagents and the results obtained from samples containing a low microbial biomass (or low quantities of successfully amplified target DNA) can be hampered by these contamination levels [30], [31]. While the low number of reads was expected in the controls (given the amount of DNA encountered after quantification) the lack of amplification within all the samples was surprising and may be related to the presence of PCR contaminants (salt is a possibility) especially within the sampled books. Salt is a known inhibitor of *in vitro* DNA amplification techniques and might have hindered our efforts to obtain the full spectra for bacterial and fungal contamination.

The molecular biology protocols for both bacteria and fungi, despite their limitations, do point to the presence of halophilic bacteria, and Archaeobacteria both reinforcing the existence of an extreme environment.

It is less likely that the environment (assessed through EDCs) would be as affected by salt contamination. Given the care presently being put on achieving proper storage conditions (temperature and relative humidity control, protection of the shelves with contaminated books), and the constant precipitation levels in Bergen (which are expected to lower the number of spores entering the archival storages) the obtained low contamination levels may, in fact, depict a low contaminated environment, as shown by both the metagenomics and the classic culturing methods used.

As for the sampled books, they show, on average, higher levels of contamination. Our results show that – as previously concluded by other studies [3] – the most common genera found in archives are the environmental fungi *Penicillium* sp., *Aspergillus* sp. and *Cladosporium* sp. Many of the species belonging to these genera are cellulolytic and monitoring is warranted to identify possible surges in quantity and, being that the case, proceed to a new round of identification. The metagenomic approach (for sample FP6B) confirmed the presence of *Penicillium* sp., *Cladosporium* sp., *Aspergillus* sp., and added *Toxicladosporium sp.* and *Stachybotris sp.* to the list of relevant genera. The highly toxic *Toxicladosporium irritans* was also found in the university library of Coimbra [6] and *Aspergillus fumigatus* was detected in several archives in Poland and Portugal [6], [7]. *Stachybotrys* sp. is a black toxic mould associated with sick building syndrome [32] and has already been found both in the air and on paper in archives in Lithuania, Italy, Spain and Colombia [3], [5], [33], [34], [35].

Concerning the results obtained from culture-based methods we should not neglect the fact that assessing the viability of pathogenic and potentially pathogenic microorganisms is crucial when considering potential health effects and, therefore, reach a detailed risk assessment and identify the most relevant risk management measures to implement [36]. Gram-negative bacteria were observed in the present analysis, and we should ponder the presence of several pathogenic bacteria that can be a threat to human health, such as *E.coli, Klebsiella pneumoniae* or *Pseudomonas aeruginosa*, just to name a few [37]. In addition, and besides the more common fungal species, in the EDC and filter paper samples, we isolated *Aspergillus* sections, all with toxigenic potential. In the filter samples we isolated *Aspergillus* section *Fumigati*, listed by WHO as of critical priority and proposed as a surrogate of harmful fungal contamination in different indoor environments [11], [38], [39].

Although species with toxigenic potential were isolated, mycotoxin contamination was low: only one filter paper sample (FP6B) presented contamination by a single mycotoxin – Sterigmatocystin, produced by *Aspergillus* sp. Nevertheless, this scenario can change due to many aspects that influence mycotoxin production such as the fungal species present and environmental conditions (e.g. temperature, humidity and availability of nutrients) [40].

Cytotoxicity was only observed in lung epithelial cells, being higher in the filter papers than in the EDCs. Since filter papers presented the highest bacterial loads and *Aspergillus* sp. prevalence, it can be hypothesized that the differences observed in cytotoxicity might be partially attributed to these contaminants. Moreover, *Fumigati* section (which was identified in filter papers but not in EDCs) is reported to have a cytotoxic effect on macrophages, due to the production of gliotoxin [41]. Besides gliotoxin, other toxins from *Aspergillus* section *Fumigati* such as trypacidin, were also reported to be cytotoxic for lung cells [42]. Several other studies corroborate that *Aspergillus* section *Fumigati* present the highest cytotoxicity among *Aspergillus* species [43], [44], [45], [46], [47]. Although not assessed in this study, the effect of particulate matter and/or volatile organic compounds cannot be excluded.

The fungal species tested for azole-based fungicides resistance, failed to grow on supplemented media, revealing no antifungal resistance to the tested compounds. However, further studies in the scope of environmental surveillance regarding antifungal resistance should be in place, including cultural heritage settings. The changing climate has boosted the spread and acquisition of fungal diseases, leading to increased dispersion of fungi [48], forcing a higher use of azole-based fungicides and triggering the acquired azole resistance and the potential pathogenic fungi for humans, as well the toxigenic potential [49].

## 5. Conclusions

Under a One Health approach, we aimed to characterize the microbial burden in the books and storage environment of the NSTA-ULB, which due to microbial contamination is being kept from the public until further treatment deems it secure for handling. This approach was pivotal to ascertaining the risk of health effects, recommending appropriate measures in terms of protective gear and disinfection, and also increasing our ability to effectively control and remediate the biodeterioration of the historic and cultural assets present in this Archive.

The analysis of the storage environment in the NSTA-ULB revealed low levels of contamination, low cytotoxicity, and no mycotoxins, which could be related to the environmental conditions attained in this setting, besides the climate conditions of Bergen.

The collection itself presented a higher contamination burden, as would be expected, given the visible (in some cases intense) microbial colonization exhibited by the collection books. Besides the most common environmental fungi *Penicillium* sp., *Aspergillus* sp. and *Cladosporium* sp., which carry biodeterioration potential, we have also identified *Toxicladosporium* sp., *Stachybotris* sp. and *Aspergillus* section *Fumigati* all with high toxigenic potential. Also, the cytotoxicity on lung cells suggests a potential health risk for staff handling the collection. These results may explain why in the clean and stable environment of the University Library of Bergen the staff continues to experience health issues. This would justify the need to eliminate microbial remains as much as possible from the affected materials.

The obtained results can, however, be underrepresenting the real scenario, as, despite the visually evident microbial colonization on part of the studied books, we did not obtain a high burden of contamination nor a great biological diversity, either through culturing or DNA/PCR based protocols. Only one of the samples showed a substantially different fungal profile from the rest of the samples and the control samples, and it was also on that same sample that a mycotoxin was identified. These results can point out potential chemical contamination of the samples with salt (salt is a known inhibitor of in vitro DNA amplification techniques), due to the salt-rich environment of the premises where the Norwegian Sea Trade archive was originally kept. These results contribute to the growing realization that cultural heritage objects can be considered extreme environments, to which the colonizing microorganisms are well adapted, and to which specific isolation and identification methods need to be applied.

**Acknowledgements and Funding**

This research was funded by the EEA Grant Fund for Bilateral Relations "Microbiological contamination in cultural heritage settings: shared experiences for better approaches" (FBR_OC2_66_NOVA.ID.FCT).
S. Sequeira and E. Pasnak gratefully acknowledge the support by the Portuguese Fundação para a Ciência e a Tecnologia (FCT/MCTES) through research grants (CEECIND/01474/2018 and UI/BD/153082/2022) and LAQV-REQUIMTE funding (https://doi.org/10.54499/LA/P/0008/2020; https://doi.org/10.54499/UIDP/50006/2020; https://doi.org/10.54499/UIDB/50006/2020).

This research was also funded by Instituto Politécnico de Lisboa, Portugal, through the Projects IPL/2023/FoodAIIEU_ESTeSL; IPL/2023/ASPRisk_ESTeSL; IPL/2023/ARAFSawmil_ESTeSL.H&TRC and by the Polish Minister of Science and Higher Education, under the program "Regional Initiative of Excellence" in 2019 - 2022 (Grant No. 008/RID/2018/19). H&TRC authors gratefully acknowledge the national funds through FCT/MCTES (UIDB/05608/2020 and UIDP/05608/2020).

C. Pinheiro acknowledges the FCT/MCTES support through CEECIND/02598/2017. C. Pinheiro, M. Penetra, I. Santos and T.Caldeira acknowledge support from UIDB/04449/2020 (https://doi.org/10.54499/UIDB/04449/2020), UIDP/04449/2020 (https://doi.org/10.54499/UIDP/04449/2020) and LA/P/0132/2020 (https://doi.org/10.54499/LA/P/0132/2020). M. Penetra acknowledges the financial support of the project "ROADMAP - Research On Antonio De Holanda Miniatures Artistic Production" (PTDC/ART-HIS/0985/2021), financed by Portuguese funds through FCT/MCTES. I. Santos acknowledges financial support to FCT-MCTES within the scope of the project UI/BD/153582/2022.